\documentclass{article}

\usepackage{arxiv}

\usepackage[utf8]{inputenc} % allow utf-8 input
\usepackage[T1]{fontenc}    % use 8-bit T1 fonts
\usepackage{hyperref}       % hyperlinks
\usepackage{url}            % simple URL typesetting
\usepackage{booktabs}       % professional-quality tables
\usepackage{amsfonts}       % blackboard math symbols
\usepackage{nicefrac}       % compact symbols for 1/2, etc.
\usepackage{microtype}      % microtypography
\usepackage{lipsum}		% Can be removed after putting your text content
\usepackage{graphicx}
\usepackage{natbib}
\usepackage{doi}
\usepackage{amsmath}
\usepackage{amssymb}

\title{Laboratory-Based Correlative Soft X-ray and Fluorescence Microscopy in an Integrated Setup}

\usepackage{authblk}

\title{Laboratory-Based Correlative Soft X-ray and Fluorescence Microscopy in an Integrated Setup}
\author[1,2]{Julius Reinhard}
\author[1]{Sophia Kaleta}
\author[1]{Johann Jakob Abel}
\author[1]{Felix Wiesner}
\author[1, 2]{Martin Wünsche}
\author[3]{Eric Seemann}
\author[4]{Martin Westermann}
\author[1]{Thomas Weber}
\author[1, 2]{Jan Nathanael}
\author[5]{Alerxander Iliou}
\author[6]{Henryk Fiedorowicz}
\author[5, 7]{Falk Hillmann}
\author[8, 9]{Christian Eggeling}
\author[1, 2]{Gerhard G. Paulus}
\author[1, 2, 10]{Silvio Fuchs}
\affil[1]{Institute of Optics and Quantum Electronics, Friedrich Schiller University Jena, Max-Wien-Platz 1, 07743 Jena, Germany}
\affil[2]{Helmholtz Institute Jena, GSI Helmholtzzentrum für Schwerionenforschung GmbH, Fraunhofer Str. 8, 07743 Jena, Germany}
\affil[3]{Institute of Biochemistry I, Jena University Hospital, Nonnenplan 2, 07743 Jena, Germany}
\affil[4]{Electron Microscopy Center, Jena University Hospital, Ziegelmühlenweg 1, 07743 Jena, Germany}
\affil[5]{Leibniz Institute for Natural Product Research and Infection Biology, Hans Knöll Institute (Leibniz- HKI), Adolf-Reichwein-Str. 23, 07745 Jena, Germany}
\affil[6]{Institute of Optoelectronics, Military University of Technology, Kaliskiego 2, 00-908 Warsaw, Poland}
\affil[7]{Biochemistry/Biotechnology, Faculty of Engineering, Hochschule Wismar University of Applied Sciences Technology, Business and Design, Philipp-Müller-Str. 14, 23966 Wismar, Germany}
\affil[8]{Leibniz Institute of Photonic Technology e.V., Albert-Einstein Strasse 9, 07745 Jena, Germany}
\affil[9]{Institute of Applied Optics and Biophysics, Friedrich Schiller University Jena, Max-Wien-Platz 1, 07743 Jena,  Germany}
\affil[10]{Laserinstitut Hochschule Mittweida, University of Applied Science Mittweida, Technikumplatz 17, 09648 Mittweida, Germany}

\renewcommand{\headeright}{}
\renewcommand{\undertitle}{}
\renewcommand{\shorttitle}{Laboratory-Based Correlative Soft X-ray and Fluorescence Microscopy in an Integrated Setup}

\hypersetup{
pdftitle={Correlative Microscopy},
pdfsubject={physics.optics, physics.ins-det},
pdfauthor={Julius Reinhard},
pdfkeywords={First keyword, Second keyword, More},
}

\begin{document}
\maketitle

\begin{abstract}
Correlative microscopy is a powerful technique that combines the advantages of multiple imaging modalities to achieve a comprehensive understanding of investigated samples. For example, fluorescence microscopy provides unique functional contrast by imaging only specifically labeled components, especially in biological samples. However, the achievable structural information on the sample in its full complexity is limited. Here, the intrinsic label-free carbon contrast of water window soft X-ray microscopy can complement fluorescence images in a correlative approach ultimately combining nanoscale structural resolution with functional contrast. However, soft X-ray microscopes are complex and elaborate, and typically require a large-scale synchrotron radiation source due to the demanding photon flux requirements. Yet, with modern high-power lasers it has become possible to generate sufficient photon flux from laser-produced plasmas, thus enabling laboratory-based setups. Here, we present a compact table-top soft X-ray microscope with an integrated epifluorescence modality for 'in-situ' correlative imaging. Samples remain in place when switching between modalities, ensuring identical measurement conditions and avoiding sample alteration or destruction. We demonstrate our new method by multimodal images of several exemplary samples ranging from nanoparticles to various multicolor labeled cell types. A structural resolution of down to 50\,nm was reached.
\end{abstract}


\section{Introduction}
Utilization of different imaging techniques to collect data from a sample allows to obtain more comprehensive information about its properties. This so-called correlative imaging is especially useful if complementary imaging techniques with different contrast mechanisms are combined. A particularly promising example is the combination of fluorescence and soft X-ray (SXR) microscopy in the water-window \citep{fonta2015correlative}. Fluorescence microscopy (FLM) is a powerful tool in its own right for examining a variety of biological samples; it is in fact one of the most popular techniques in the life sciences \citep{lichtman2005fluorescence} especially with the rise of super-resolution techniques \citep{schermelleh2019super}. All FLM techniques are based on labeling specific components of the sample with fluorescent markers. This provides excellent functional contrast. But the greatest advantage of the method is also its greatest disadvantage: The sample typically cannot be imaged in its entirety as only labeled components are visible. This gap can adequately be filled by using soft X-ray (SXR) microscopy in the so called water window as a complementary correlative method. In the water window (WW) spectral range, which is defined by the absorption edges of carbon (282\,eV/4.4\,nm) and oxygen (533\,eV/2.3\,nm), a strong and label-free structural contrast can be achieved for biological samples with a resolution of down to 10\,nm, while still offering a relatively high penetration depth of several micrometers into water. A wide variety of imaging techniques have been established in this energy range, examples being coherent diffraction imaging (CDI), ptychography \citep{chapman2010coherent, rose2018quantitative}, X-ray holography \citep{mancuso2010coherent} and Fresnel zone plate (ZP)-based methods \citep{jacobsen2019zone} such as scanning transmission X-ray microscopy (STXM) \citep{chao2012real} or wide field imaging \citep{legall2012compact}. Due to the relatively high penetration in water even tomography is possible \citep{schneider2010three}. At synchrotron facilities also correlative FLM-SXR microscopy has been demonstrated \citep{bernhardt2018correlative, duke2014imaging, smith2014quantitatively, hagen2012correlative, varsano2016development}. In all of these examples, fluorescence microscopy was used to identify and/or image the cellular components relevant to the research question. The structural and label-free contrast of SXR microscopy combined with its high resolution then allowed these components to be viewed in the context of their environment, providing additional information. A comparable contrast may be achieved through different techniques of phase-contrast microscopy \citep{zernike1935phase, lang1982nomarski}. However, SXR microscopy allows much smaller structures to be identified.

Many of the above-mentioned SXR methods are actually limited to synchrotron radiation sources, as they require a high flux of coherent photons, which cannot yet be generated in the laboratory. However, since access to these large-scale facilities is limited, it is of great interest to endow laboratory-scale SXR microscopy with the FLM modality to maximize the impact of both methods. 

Generating sufficient photon flux in the WW spectral region is a challenge for laboratory-based setups. While WW coherent sources based on high harmonic generation exist \citep{gebhardt2021bright} their flux is still way too low for studies of biological samples. For this reason, most laboratory SXR microscopes are driven by incoherent plasma sources, where laser-produced plasmas have proved being the most powerful. Various target materials can be used ranging from solids \citep{fahy2021compact} and (cryogenic) liquids \citep{berglund1998cryogenic, berglund2000compact} to gas targets \citep{muller2014table,wachulak2015desktop}. A detailed overview on the development of laboratory water-window microscopy is given in \citep{kordel2020laboratory}. Although a number of laboratory-based SXR microscopes exist, to our knowledge a correlative instrument consisting of a compact SXR and FLM has not been demonstrated. What has been demonstrated is the combination of SXR with conventional light microscopy to accelerate the tomography measurement routine \citep{dehlinger2020laboratory}.

\begin{figure*}[!t]%
\centering
\includegraphics[width=\textwidth]{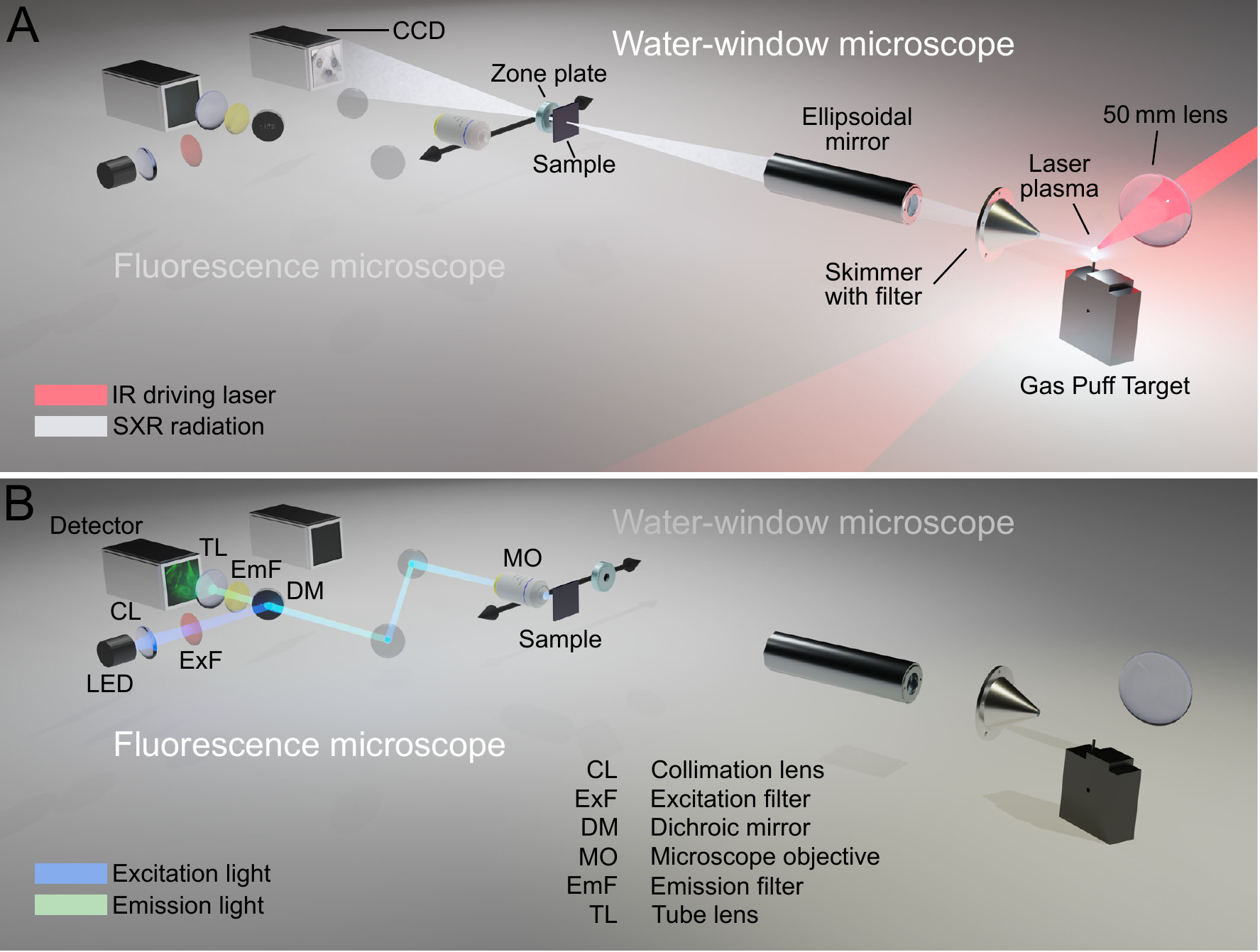}
\caption[Sketch of full setup]{Sketch of the full setup. \textbf{A)} The top figure shows the running SXR microscope with the ZP imaging the sample structure on the detector. The fluorescence microscope is not active in this mode. \textbf{B)} The zone plate, objective and the first mirror behind it are moved sideways to allow the fluorescence microscope to image the sample, which remains in place. The laser and plasma source are not active in this mode.}
\label{fig:setup}
\end{figure*}

In this work, we present a compact table-top SXR microscope with an integrated epifluorescence modality for ”in-situ” correlative imaging. A sketch of our setup is shown in Figure \ref{fig:setup}. Nitrogen gas is utilized as the target for the generation of monochromatic SXR line emission at a wavelength of 2.88\,nm. Using a tube-shaped, axisymmetric elliptical mirror as condenser optics \citep{hudec2000replicated}, a wide-field zone plate microscope was built allowing for a structural resolution of 50\,nm. Conventionally, correlative imaging is achieved by moving the sample from one microscope to another, oftentimes with additional preparation steps in between \citep{fonta2015correlative}. Here, we directly integrated an epifluorescence microscope with different filter sets into the SXR microscope. This allows 'in-situ' correlation, i.e. there is no need to move or remove the sample from the SXR microscope's sample holder when switching between the different modalities.

Besides the obvious benefit of combining the functional fluorescence contrast with the natural structural contrast of WW microscopy, this offers several additional advantages: The field of view (FOV) of the SXR microscope is relatively small ($<$60\,\textmu m), which complicates the identification of regions of interest, which is particularly relevant in consideration of the long exposure times in the SXR. By integrating the FLM into the SXR setup, it is possible to scan large areas of the sample quickly with the FLM to easily find regions of interest. In addition, the two images can be taken immediately after each other, so changes of the sample can be avoided. We demonstrate this with exemplary samples ranging from fluorescent nanoparticles and cyanobacteria to labeled 3T3 and COS-7 cells. These examples highlight how FLM can guide SXR observation to regions of interest or associate signatures in SXR images to specific cellular structures. Furthermore, our setup also allows for the investigation of the fluorescence response under SXR illumination to be studied, which is of particular interest for future applications. Our synchrotron-independent compact device, with a footprint of only 1.5\,m\,$\times$\,4\,m, has the potential of becoming a stand-alone tool in biological research labs adding the unique label-free SXR structural contrast to the manifold of imaging methods.

\section{Materials and Methods}
\subsection{SXR Microscope}
The SXR microscope has the same basic design as a wide-field optical microscope. A laser-generated plasma is used as an X-ray source. As a condenser serves an ellipsoidal mirror for illuminating the specimen. The objective of the microscope is a Fresnel zone plate, which images the sample onto a CCD camera. The full setup including the fluorescence microscope is shown in Figure \ref{fig:setup}. Due to the strong absorption of soft X-rays in air, the microscope is operated in a vacuum. In the following sections, the individual components are described in detail.

\subsubsection{Plasma Source}
The SXR radiation is generated in a laser-produced plasma utilizing nitrogen as target gas. While it is possible to use solid or liquid targets for higher plasma densities, a gas target has the advantage of producing very little debris. In addition, it is technically easy to implement. The gas-nozzle used is a so called double-stream gas puff target (GPT) \citep{wachulak2010water}, which has been employed for various applications requiring extreme ultraviolet or SXR radiation ranging from microscopy \citep{wachulak2015desktop} to XUV coherence tomography \citep{fuchs2017optical, wachulak2018optical, skruszewicz2021coherence}. It consists of two circular concentric nozzles. The actual target gas streams out of the inner nozzle, while a low absorption, low Z-number gas (typically helium) is emitted from the outer nozzle. The latter limits the target gas expansion, thus allowing higher densities even at larger distances from the nozzle, which leads to higher photon flux. Nitrogen is used as the working gas because it provides isolated emission lines in the water window region at a favorable plasma temperature of about 150\,eV \citep{NIST_ASD}. In order to reach the required plasma conditions, a commercial Nd:YAG laser system (Spectra-Physics, Quanta Ray Pro-350) is tightly focused with a 50-mm aspheric lens into the gas stream above the nozzle. The laser pulses have an energy of up to 2.5\,J and a pulse duration of 10\,ns at 1064\,nm wavelength and 10\,Hz repetition rate.

\subsubsection{Condenser Optics}
The emitted SXR-radiation is collected and focused by a nickel-coated ellipsoidal mirror (Rigaku) with a distance of 400\,mm between the two focal spots. The ellipsoid has a length of 105\,mm, an input NA of 0.05 to 0.1 and a focusing NA of 0.03 to 0.05. As a consequence, the reflection angles are 3$^\circ$, resulting in a reflectivity as high as 76\%.

In the setup, the optical axis is oriented perpendicular to the driving laser, as shown in Figure \ref{fig:setup}. For proper alignment of the mirror, the adjustment of tip, tilt, and translation is possible in all directions. In the respective procedure, the unfocused annular beam is viewed by the in-vacuum camera near, but not at the focal point and adjusted for maximum symmetry. The same camera was also used to characterize the resulting focal spot. To prevent the camera from becoming oversaturated, an additional, approximately 4\,\textmu m of aluminum foil was placed behind the condenser to mitigate the light. The focal spot is presented in Figure \ref{fig:plasma_charac}.C. The image is slightly smoothed to filter out the effects of the uneven filter foil. The focus is approximately Gaussian with a FWHM width and height of 750\,\textmu m\,$\times$\,675\,\textmu m, which corresponds well to the previously determined size of the plasma itself.

An axisymmetric ellipsoidal mirror has a single stigmatic pair, comprising the two focal points. This results in solely the light emanating from one of the focal points converging onto the other. Rays emitted from any other position, regardless of whether they are shifted in the axial or lateral direction, will form an annular shape in the focal spot plane. As such, there is no magnified or shrunken image of the plasma and no magnification can be defined. Therefore, the focal spot is relatively round and symmetrical even if the source is oval with an aspect ratio of 2:1. Nevertheless, the FWHM of the focus is slightly increased as compared to the source size due to the mirror being closer to the source than the sample. This behavior was expected and confirmed by simulations with the ray-tracing software OpticStudio.

\subsubsection{Objective and Detector}
Because of the high absorption of almost all materials in the WW spectral region, it is not possible to use refractive optics such as glass lenses or objectives for imaging. Instead, a Fresnel zone plate is employed, which uses diffraction instead of refraction for image formation. At a certain distance, a focus is created by constructive interference. The theoretical resolution is determined by the outermost zone-width $\Delta r_N$: $d_\text{Rayl}=1.22\Delta r_N$ \citep{attwood2016x}.

We use a ZP with a diameter of 180\,\textmu m and an outer-most zone width of 33\,nm. This results in a focal length of 2.06\,mm at 431\,eV photon energy and offers a theoretical resolution of 40\,nm. The NA of 0.044 is matched to the condenser NA, providing incoherent illumination for maximum contrast \citep{heck1998resolution}. The ZP was manufactured on a silicon nitride (Si$_3$N$_4$, short SiN) membrane in 150\,nm tungsten by Zoneplates Ltd. The image is detected with a back-illuminated CCD camera (Andor iKon-L, 2048$\times$2048 pixels, 13.5\,\textmu m pixel size), which can be mounted at distances varying from 500\,mm to 1000\,mm from the ZP, depending on the desired magnification (250 to 500).

\subsubsection{Pumping}
A differential pumping scheme is employed to reduce the pressure in the main experimental chamber, where all measurements take place. At the position of the skimmer and filter a partition was installed between the source and measurement chambers, such that the GPT resides in its own small vacuum chamber. The latter is pumped by a roots-pump (Edwards iXL 600) to keep the pressure at roughly $10^{-1}$\,mbar during measurements. The relatively high pressure is almost exclusively due to the difficult-to-pump helium from the larger outer nozzle of the GPT. Accordingly reabsorption of generated SXR radiation is nevertheless low. The main chamber is pumped by two turbomolecular pumps (Pfeiffer Vacuum HiPace), resulting in a pressure of about $10^{-4}$\,mbar when the gas nozzle is running. Both turbomolecular pumps are backed by a scroll-pump (Pfeiffer Vacuum HiScroll 18).

\begin{figure*}[!t]
	\centering
	\includegraphics[width=\textwidth]{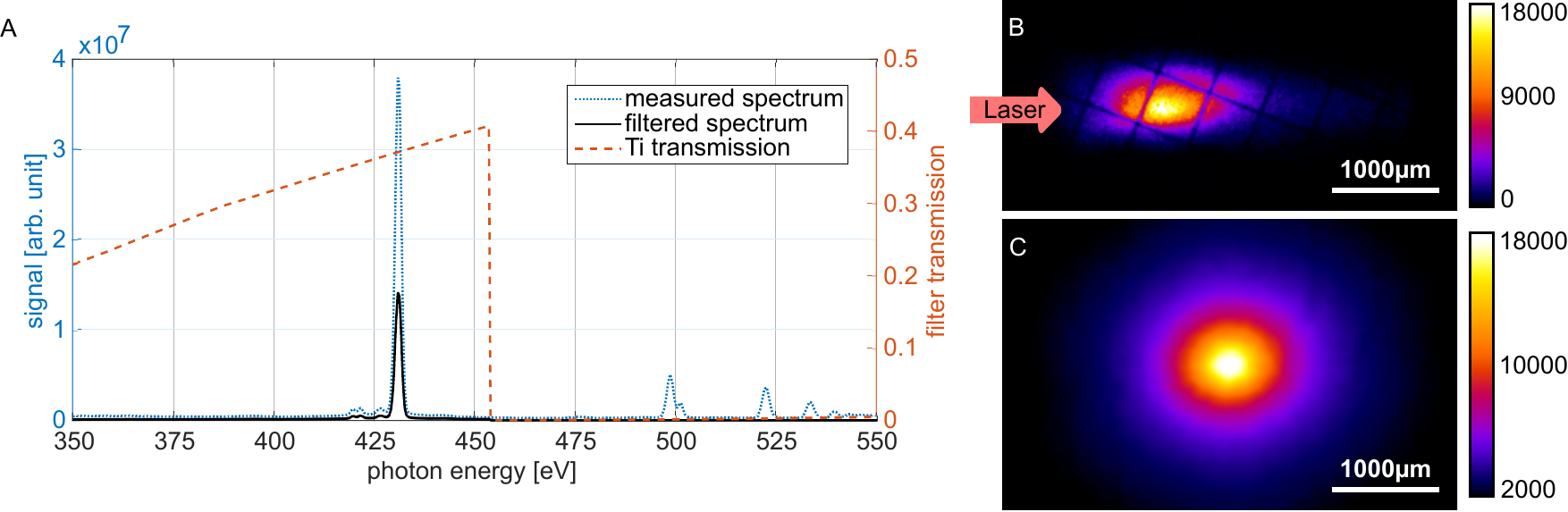}
	\caption[Plasma characterization]{Spectral and spatial characterization of the SXR plasma source. \textbf{A)} The measured spectrum shows a strong line emission at 431eV and multiple lines at higher energies. By using a 600\,nm Titanium foil the spectrum can be monochromatized. \textbf{B)} Direct image of the source recorded by a pinhole camera with titanium filters attached. The FWHM diameter of the plasma is 325\,\textmu m$\times$675\,\textmu m. \textbf{C)} Image of the condenser focus. It is almost circular and has a FWHM diameter of 675\,\textmu m$\times$750\,\textmu m.}
	\label{fig:plasma_charac}
\end{figure*}

\subsection{Fluorescence Microscope}
For correlative imaging, a fluorescence microscope (FLM) was integrated into the SXR microscope in the form of a bright-field epi-illuminated setup \citep{Webb2013}. The scheme is shown in Figure \ref{fig:setup}.B. It has been realized in such a way that only the objective and two mirrors are located inside the vacuum chamber. All other parts of the FLM are outside.

The excitation light from a fiber-coupled LED is collected by a 30\,mm achromatic lens and guided through a spectral filter to a dichroic mirror, which reflects the light towards the vacuum chamber. Inside the chamber the light passes through the microscope objective (Olympus UPlanSapo 40x2) and illuminates the sample. The fluorescent light emitted by the sample is collected by the same objective and directed back as a collimated beam. Due to the Stokes shift, it has a longer wavelength than the excitation light and thus can be transmitted by the dichroic mirror. Additional filtering is used to mitigate the background signal from excitation light reflected at the sample. A tube lens is then used to create the fluorescence image on the detector (pco edge). Leaving all components except the lens in air allows easy switching between different filter sets and illumination LEDs. This has the decisive advantage of enabling multi-color fluorescence imaging. The objective is motorized for translation in all dimensions such that alignment and refocusing is always possible.

The objective of the FLM is placed next to the ZP on a shared stage. This allows fast switching of the imaging modality without the need to move the sample. The mirror directly behind the objective is moved together with the optics such that the SXR beam behind the ZP can reach the detector unobstructed.

Three different filter sets (Chroma Technology) were used so far: One for UV/blue fluorescence (emission window: 375\,nm central wavelength, 28\,nm bandwidth, cutoff wavelength of dichroic mirror 415\,nm, excitation window: 460\,nm central wavelength, 50\,nm bandwidth), one for green (480/40, 510, 535/50) and one in the deep red region (615/40, 635, $>$638). These filters were used in combination with different fiber-coupled LEDs from Pyroistech (LEDM-365, 365\,nm), Mightex (FCS-0490, 490\,nm) and Thorlabs (M625F2, 625\,nm).

The integrated microscope enables seamless switching between modalities without moving the sample, ensuring that the captured images display the same area of the sample (excluding the different size of the FOVs). The precise correlation is subsequently performed semi-automatically by identifying three prominent features visible in both methods. The rotation and scaling factors between both methods are determined by comparing the path lengths and angles of the lines connecting these features.

\subsubsection{Photobleaching by SXR Irradiation}
It is worth noting that the setup can be used to quantitatively study the behavior of fluorescent dyes under SXR irradiation. These investigations can be of particular interest in view of the increasing use of correlative imaging methods, be it with X-ray or with electron microscopy. For respective investigations, the objective is moved to the "fluorescence mode" in front of the sample and used with the appropriate LED. At the same time, the sample is irradiated from behind using the laser and the plasma source. By taking frequent fluorescence images during this process, the decay of the signal due to the ionization and destruction of the fluorophores by the X-rays can be measured. Due to the preparation of the samples on SiN membranes, this measurement can even be carried out with a reference. To this end, particles on the free-standing SiN membrane and the silicon substrate next to the membrane are imaged simultaneously with the fluorescence microscope. Since the SXR radiation does not penetrate the 200\,\textmu m thick silicon, these particles are not affected.

\subsection{Samples and Preparation}
The samples to be investigated were chosen to demonstrate the capabilities of the microscope with increasing complexity. In order to characterize the resolution of the SXR microscope a Siemens star with a diameter of 60\,\textmu m and structure sizes decreasing to 50\,nm half pitch towards the center of the star was investigated. This sample has been manufactured in a similar way to the ZP via electron beam lithography on a SiN membrane in 175\,nm tungsten.
All the other samples were prepared on SiN membranes (Norcada Inc). These membranes offer several advantages, namely transparency to both X-rays and visible light, mechanical stability, and compatibility with different sample preparation methods. For this reason, they have been used for SXR microscopy since the early development of the technique \citep{anastasi1992preparation}. They are manufactured on 5\,mm$\times$5\,mm silicon wafers and typically have a thickness around 50\,nm. The silicon is etched away such that a free-standing SiN window with a size of up to a few mm remains. For our samples we typically use window sizes between 150\,\textmu m\,$\times$\,150\,\textmu m and \,500\textmu m\,$\times$\,500\,\textmu m, as they provide a good compromise between open aperture and stability.

Since X-ray microscopy requires a vacuum environment, there are also special requirements for the specimens or holders. Therefore, only dry samples were examined so far. For this purpose, samples were dried either in air or in vacuum (nanoparticles and cyanobacteria) or subjected to critical point drying (NIH3T3- and COS7-cells). This allows easy handling and ensures sufficient radiation resistance. Investigation of wet samples presents greater challenges and will be addressed in future research.

The first samples for demonstrating the correlation of a strong fluorescence signal and good SXR contrast were three different types of fluorescent nanobeads (FluoSphere Carboxylate-Modified Microspheres, excitation/ emission 360/ 450, 480/ 520 and 625/ 645) with sizes of 1\,\textmu m and 200\,nm. They were investigated with the three different filter sets and LEDs described above. For the preparation on the membranes the bead suspensions were diluted in water in a ratio of 1:1000 for the 200\,nm beads and 1:500 for the 1\,\textmu m beads according to the results of some preliminary tests on glass slides. The initial density of the bead solutions before mixing with water was 2 percent. To achieve sufficient dispersion of the beads and to minimize clustering, the water-mixed beads were placed in an ultrasonic bath for five minutes. Then the membranes were immersed in poly-L-lysine for 15 minutes before being washed with distilled water in order to ensure a high particle adherence while avoiding clustering. After that, the particles were applied and pipetted off after 10 minutes. This led to a sufficient density of approximately uniformly distributed particles on the membrane with only a few clusters.

The first biological samples investigated were cyanobacteria of the type Synechocystis sp. PCC6803. These bacteria are autofluorescent due to their chlorophyll content. In addition, they are sufficiently robust to limit structural changes under vacuum conditions. The excitation of chlorophyll is possible over a broad spectral range. Therefore the red filter set was used. The bacteria were grown under constant light illumination until an optical density of 1.5 was reached at $\lambda=$720\,nm. Then, the cells were harvested by centrifugation, washed with H$_2$O and resuspended in H$_2$O. Different dilutions were dropped on poly-L-lysin coated SiN membranes. Cells were airdried and stored at 4°C until imaging. 

For a demonstration of the capabilities and scientific potential of the correlative microscope, a conventional cell culture with multiple fluorescent labels was investigated. To this end, NIH-3T3 and COS-7 cells (Cell Lines Services GmbH) were cultured as described in \citep{seemann2017deciphering}. The cells were seeded and grown on SiN membranes. After the cells reached a confluence of 75\%, they were incubated for 10\,min with Mitotracker Deep Red 633 at 37$^\circ$C and then fixed with 4\% PFA for another 10 min. Immunofluorescence staining was done according to \citep{schneider2014prosap1}. After quenching with 25\,mM glycine in PBS for 30\,min, the cells were permeabilized and blocked with 10\% horse serum and 2\% BSA in PBS (blocking solution) with 0.2\% Triton X-100. Alexa Fluor 488 phalloidin and DAPI incubations were done in blocking solution for 1\,h at room temperature with PBS washing steps. The cells were stored in 4\% PFA. Prior to critical-point drying in a Leica EM CPD300 automatic critical point dryer, the samples were dehydrated in ascending ethanol concentration (30, 50, 70, 90, 100\%) for 10\,min at each concentration.

\subsubsection{Sample Holder}
In order to examine new specimens, the vacuum chamber must be ventilated and evacuated for the transfer, causing the focus positions of the optics to shift slightly. It is therefore useful to be able to place as many samples as possible in the microscope at the same time. As all samples are prepared on SiN membranes of the same size, a sample holder was developed that can hold up to 33 SiN membranes and the Siemens star. It is mounted on a 2D-translation system from SmarAct, which allows precise lateral movement perpendicular to the optical axis over a range of 100\,mm\,$\times$\,100\,mm. In addition, the distances between the various membranes on the sample holder are known, which greatly speeds up the search for the exact sample position.

\section{Results and Discussion}
\subsection{Plasma Characterization}
In order to validate the emitted radiation, a spectrometer consisting of a 2400-lines/mm VLS-grating (Hitachi) and a CCD-Camera (Andor Newton) was set up. The design of the device is based on the one of \citep{wunsche2019high}, but has been optimized for the present application. Respective measurements are shown in Figure \ref{fig:plasma_charac}.A. Dominant line emission from the 1s$^2$-1s2p-transition in He-like nitrogen at 431\,eV is visible with additional lines at higher energies. Due to the large source size, the spectral resolution of the spectrometer is limited. We assume that the spectral line width is in fact much smaller than indicated here. Since monochromatic illumination is required for ZP-microscopy, two 300-nm Titanium absorption filters were placed between source and condenser. Two filters are necessary, because a single filter would still transmit some visible light through micro holes. The transmission curve (red) of 600\,nm titanium is shown in Fig. \ref{fig:plasma_charac}.A \citep{henke1993x}, along with the monochromatized spectrum calculated from the measured spectrum and the filter transmission (41\% @\,431\,eV). Taking into account the aperture and efficiency of the grating as well as the camera efficiency and transmission of the used filters, a photon flux of $\approx3\times10^{11}$ photons/(sr$\times$pulse) was calculated. This compares quite well to a similar microscopy setup with a GPT \citep{wachulak2015desktop}. The SXR microscope of the Stockholm group, which reported the highest photon flux for a laboratory based WW microscope to date, reaches $5.5\times10^{11}$ photons/(sr$\times$pulse) \citep{martz2012high}. However, it is based on a cryogenic nitrogen source working at 2\,kHz instead of 10\,Hz.

The size of the generated plasma has been measured using a pinhole camera setup. For this purpose a 20\,-\textmu m pinhole was placed 20\,mm behind the source. Another 20\,mm behind the pinhole an in-vacuum CCD camera (greateyes GE-VAC) was placed to record the image as shown in Figure \ref{fig:plasma_charac}.B. In this case, a single titanium filter was placed between source and detector so that only the hot and dense plasma emitting the 431\,eV line is imaged. The support mesh of the filter is visible in the recorded image. Moreover, it can be observed that the plasma is neither horizontally nor vertically symmetrical. This can be explained as follows: The gas was hit by the laser on the left side as indicated by the red arrow. Due to the gradual absorption of the laser during propagation through the plasma, the highest intensity is observed slightly shifted to the left. In addition, self-focusing in the plasma causes a prolonged tail of lower intensity on the right side, resulting in a horizontal extension of 700\,\textmu m FWHM. In vertical direction, the maximum intensity is slightly shifted downward, which is a direct consequence of the higher gas density closer to the nozzle. The vertical FWHM is just 340\,\textmu m. The difference in size originates from the focusing geometry.

\subsection{Siemens Star}
For characterization of the X-ray microscope, a Siemens star resolution test target was examined as a first step. This is a circular object with a diameter of 60\,\textmu m, comprising of 29 rings. The rings have a width of 1\,\textmu m and consist of transmitting and absorbing sectors. The sector spacing reduces towards the center of the circle, with the smallest structure size being found in the innermost ring, at a bar width of 50\,nm. With each successive ring, the structure size increases by 50\,nm.

The results of this investigation are shown in Figure \ref{fig:Siemens}. In \ref{fig:Siemens}.A a 80\,s exposure showing the entire sample is presented. The field of view (FOV) is about 50\,\textmu m in each direction. The red inset indicates the area, which is enlarged in subfigures B to E. In the 80\,s exposure, the second ring from the center can be clearly resolved with 100\,nm half-pitch features and even the 50\,nm-structures of the innermost ring are recognizable (Fig. \ref{fig:Siemens}.B). 

For further improvement of the resolution and signal-to-noise ratio (SNR), the exposure time needs to be increased. This is done by stacking several successively taken images. However, simply superimposing five exposures of 80\,s each actually leads to a reduction of  resolution  as shown in Figure \ref{fig:Siemens}.C. The innermost ring is not resolvable and even the contrast on the top right and bottom left of the second ring is clearly reduced. This is due to changes in position of the image on the camera. The underlying issue is a changing temperature of the instrument caused by the exhaust heat of the laser and the pumps, which causes thermal drift of the sample, the ZP, and/or the camera.

In Figure \ref{fig:Siemens}.D, the same five images were added, however the drift was corrected by post-processing.  The respective algorithm will be described in the following section. It allows for a significant increase in resolution: Stacking only five exposures, equivalent to an exposure time of 400\,s, the structures in the innermost ring can be resolved, indicating a resolution of 50\,nm half-pitch (Fig. \ref{fig:Siemens}.D). Additional exposures increase the SNR even more, as shown in Figure \ref{fig:Siemens}.E, where the innermost ring and its structures are clearly visible.

\subsubsection{Drift Correction}
\begin{figure*}[!t]
    \centering
	\includegraphics[width=\textwidth]{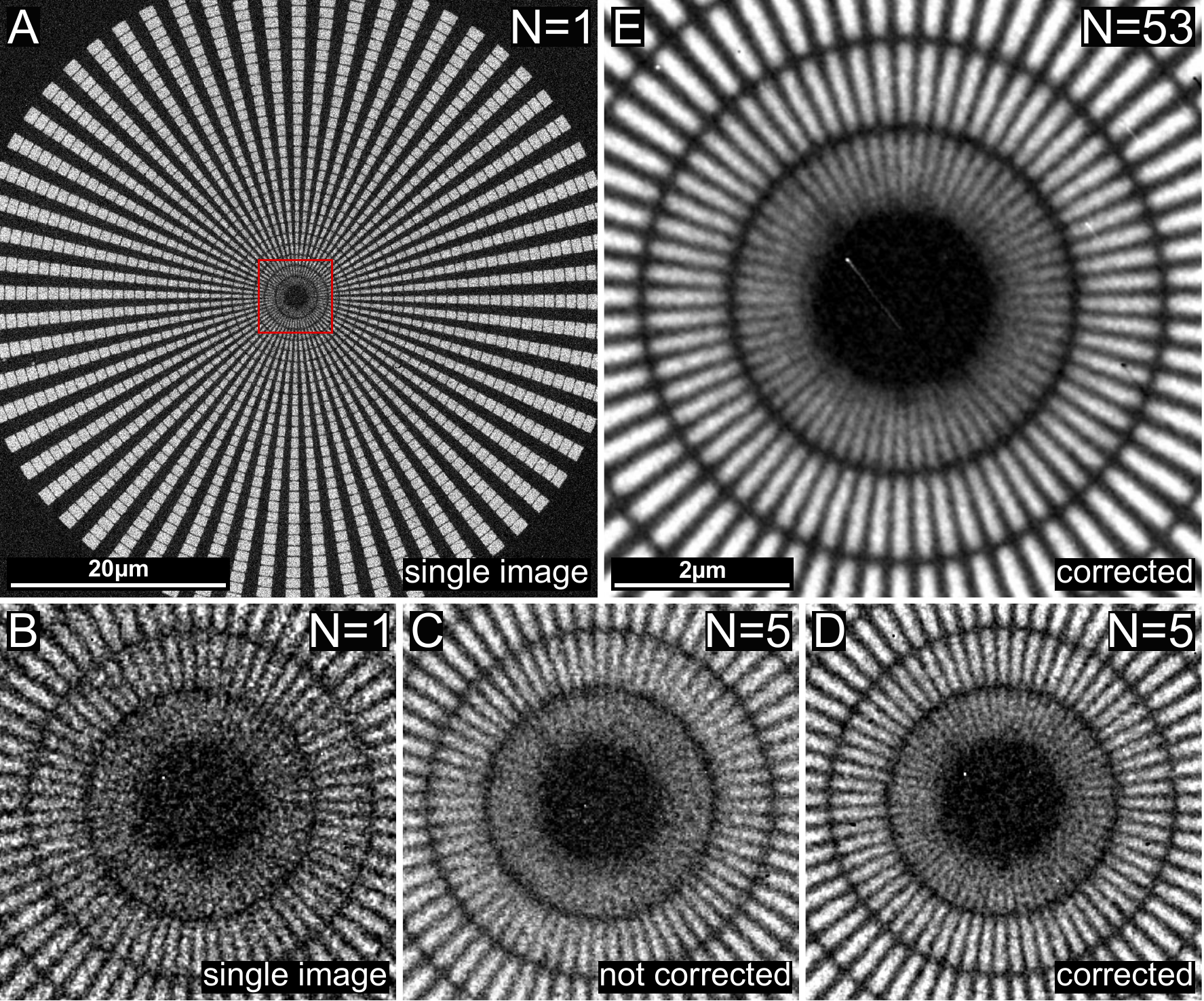}
	\caption[Siemens star]{SXR images of a Siemens star resolution target consisting of rings with structures decreasing in size towards the center. The innermost ring has structure sizes of 50\,nm, the second ring of 100\,nm. \textbf{A)} Single 80\,s exposure, full FOV. \textbf{B)} Single image, zoom on red inset, innermost structures. \textbf{C)} 5 images added up without drift correction. \textbf{D)} 5 images added up with drift correction. \textbf{E)} 53\,$\times$\,80\,s exposure for maximum contrast.}
	\label{fig:Siemens}
\end{figure*}

\begin{figure*}[!t]
    \centering
	\includegraphics[width=\textwidth]{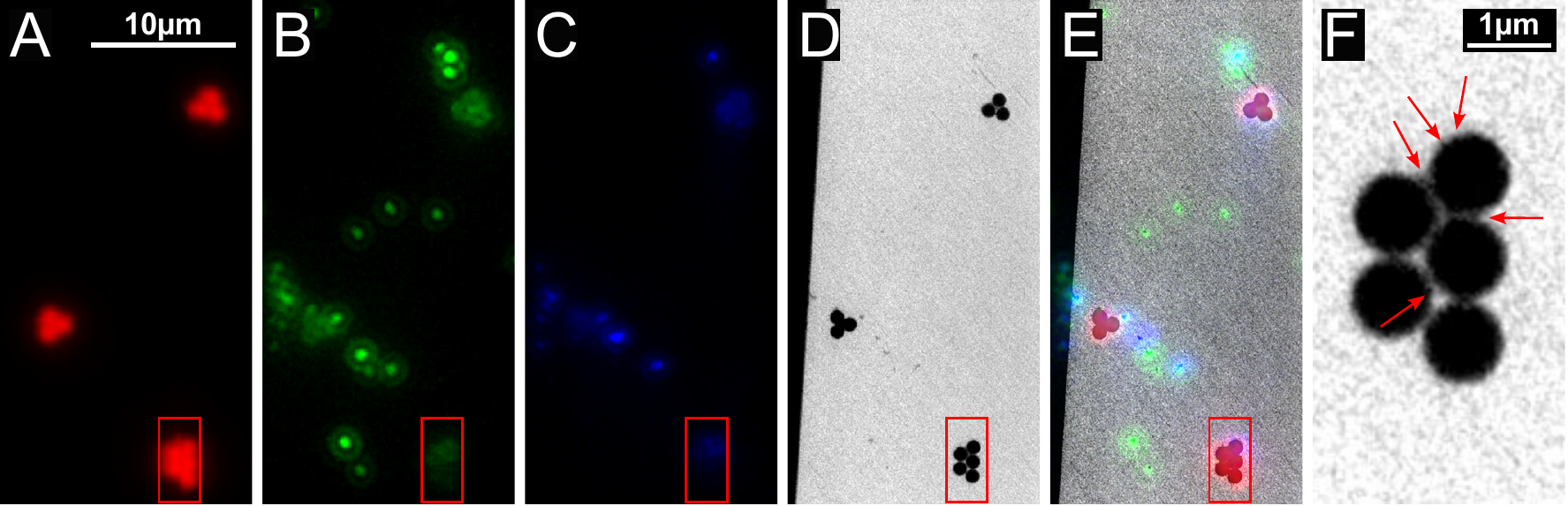}
		\caption[Correaltive Nanobeads]{Correlative measurement of fluorescent nanobeads on SiN membrane. \textbf{A)} 1\,\textmu m red fluorescent beads. \textbf{B)} 200\,nm green fluorescent beads. \textbf{C)} 200\,nm blue fluorescent beads. \textbf{D)} 90\,$\times$\,45\,s SXR image shows 1\,\textmu m beads with strong contrast and small beads scattered around. \textbf{E)} Composite image of all 4 channels, some crosstalk between different colors is visible. \textbf{F)} Enlarged image of the red frame, showing 5 large beads and some small beads attached to them.}
	\label{fig:beads}
\end{figure*}

Due to the continuous thermal drift of the sample relative to the camera position, it is not possible to take a single long exposure. Instead, several short exposures are summed up after their shift has been corrected. The determination of the changes in position of the individual images on the camera is the prerequisite for drift correction. To this end, a cross-correlation (CC) of all exposures with a specific reference image is calculated. From the position of the correlation maximum, the displacement can be determined. For computing the CC, we do not use the entire image, but rather a region of interest (ROI) with distinctive features and high contrast. A welcome side effect is that the computation time for the CC decreases. On the other hand, the precision of the shift measurement can be increased by interpolating the images to twice their size such that the exposures are stacked with half-pixel accuracy. For a series of 53 images of the Siemens star, each with an exposure time of 80\,s (Fig. \ref{fig:Siemens}), a drift of 770\,nm in vertical and 580\,nm in horizontal direction was detected and compensated for. The maximum drift speed for this measurement was 35\,nm/exposure, i.e. the resolution of a single image was not limited by drift. Shorter exposure times would certainly reduce the drift between subsequent exposures. However, the SNR would degrade at the same time, which would reduce the accuracy of the drift correction and consequently also the resolution. Therefore, there is an optimum for the exposure time. As a final step, the drift curve obtained is smoothed under the assumption that the thermal drift is more or less linear and  discontinuities are not expected. Then, the drift curve is used to adjust the positions for all the 80\,s exposures.

The procedure is hampered by residue on the camera chip, which results in small features in the images that do not change position during warm-up. For this reason, the reference image is also displaced, in fact by a relatively large margin ($\approx$\,2\,\textmu m), such that the sample structure dominates the correlation result and not the residue. For biological samples like the COS-7 and 3T3 cells, a Gaussian filter is additionally used to smooth the image before correlation. This reduces not only the effects of the residue but also of noisy signal. The correlation of large structures such as the cell nuclei nevertheless provides precise information about the displacement, since it is basically a comparison of the centers-of-mass of the images, i.e. sharp edges are not required.

\subsection{Fluorescent Nanoparticles}
The correlative imaging performance  of the microscope was tested using three different types of fluorescent nanoparticles, which had been  prepared on a SiN membrane. The red fluorescent beads have a diameter of 1\,\textmu m, the blue and green beads 200\,nm. A few typical microscope images are shown in Figure \ref{fig:beads}. In \ref{fig:beads}.A, three clusters consisting of 3 to 5 particles each are visible, which are also evident in the SXR image (Fig.~\ref{fig:beads}.D) with a high contrast.

A FLM can detect structures that are well below the resolution limit, as long as they are fluorescent. Accordingly, also the small green (Fig.~\ref{fig:beads}.B) and blue (Fig.~\ref{fig:beads}.C) beads, which are dispersed more widely across the image, are easy to recognize. The SXR microscope, on the other hand, can only image structures above the resolution limit. The further the structures are above the resolution limit, the stronger the contrast. This behavior is generally described by the modulation transfer function. Accordingly, the blue and green particles, which are only 200\,nm in size, are much more difficult to detect. In addition, due to their small thickness, the absorption of the particles is also small, further reducing the contrast. Nevertheless, all of them can be recognized by close inspection of Figure \ref{fig:beads}.D. It should be mentioned that it would be hard to distinguish the small beads from  residue on the camera. Interestingly, thermal drift and its subsequent compensation are helpful in this case, because the drift compensation procedure smeares residue into lines, while the particles remain dots. 

Because of the mixing of the particles before application, the different colors can cluster together, which is clearly visible in the composite image \ref{fig:beads}.E. Blue and green are often seen in the same position, especially around the large red beads. This can also be seen by comparing Fig.~\ref{fig:beads}.B and \ref{fig:beads}.C. However, the blue and green fluorescence signals appear somewhat 'shadowed' around the large red beads (e.g. in the red frame). It is reasonable to assume that the small beads attach to the large beads, causing some of their signal to be blocked. This explanation is supported by the zoomed image \ref{fig:beads}.F (red frame in \ref{fig:beads}.A-D), where it can be seen that the large particles have small bumps on their surface, indicated by the red arrows in Fig.~\ref{fig:beads}.F. These bumps are presumably the blue and green beads. Additionally, in the gaps between the large beads multiple small structures are visible, which are probably also smaller particles. These images already show the advantage of a correlative microscope, where, on the one hand, different fluorescent nanoparticles can be identified by their color, but, on the other hand, are all visible at higher resolution in the zone plate image.

\subsection{Cyanobacteria Synechocystis sp. PCC6803}
\begin{figure*}[!t]
    \centering
	\includegraphics[width=\textwidth]{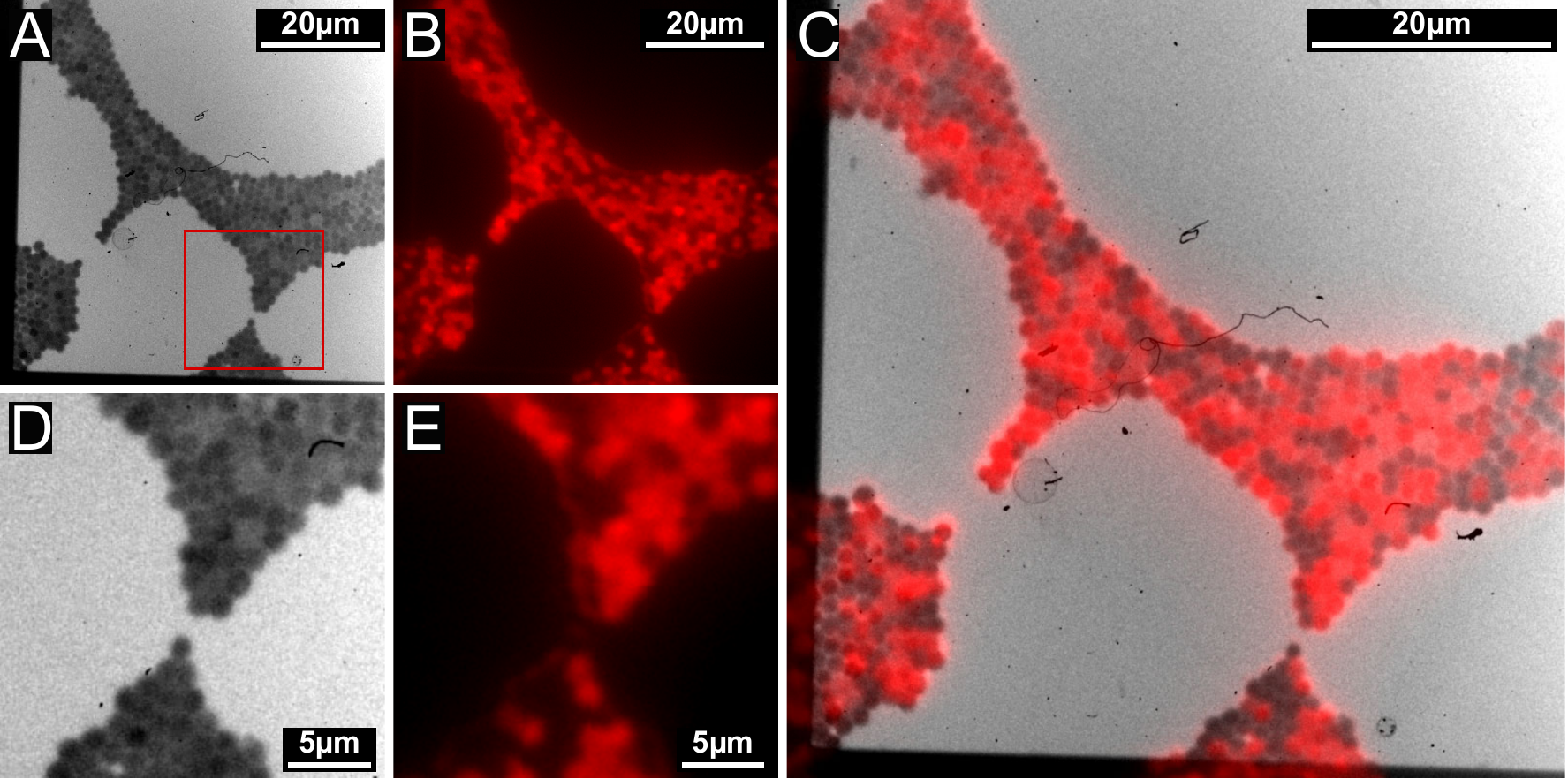}
	\caption[Cyanobacteria demonstration]{Autofluorescent cyanobacteria imaged with the red filter set (Ex/Em 625/638) and SXR radiation. \textbf{A)} SXR image after 720\,s exposure \textbf{B)} fluorescence image, full FOV. \textbf{C)} Correlative composite image of SXR and FLM. \textbf{D/E)} Enlargement of the red-framed area of panel for SXR and FLM respectively.}
	\label{fig:cyano}
\end{figure*}

The goal of our correlative microscope is its use in biological applications. Therefore, Cyanobacteria were chosen as a biological sample due their chlorophyll based autofluoresence, which can be made visible with the red filter set. The resulting images are presented in Figure \ref{fig:cyano}. The images of the full FOV (Figs.~\ref{fig:cyano}.A-C) show cluster formation of the bacteria, which originates from airdrying. In the SXR images (Figs.~\ref{fig:cyano}.A and D) individual bacteria are clearly discernible with a high resolution. The exopolysaccharide capsules of the cells are visible as small gaps between them. Different (carbon) densities of the cyanobacteria lead to differences in SXR absorption and therefore to a different contrast of each bacterium in the image. On the left and bottom side of the image, the edge of the sample membrane can be seen, as well as a small dirt fiber in the center of the image.

In comparison, the fluorescence image exhibits a different contrast originating from the uneven intracellular distribution of the chlorophyll in the bacteria. This seems to correlate with the carbon density, which can be seen when comparing Figures \ref{fig:cyano}.D and E. The darkest bacteria in the SXR image match the brightest bacteria in the fluorescence image. In addition, structures with no fluorescence signal can be identified as residue of the SXR camera or edges of the sample membrane. These examples show how the different contrast mechanisms of the two modalities nicely complement each other.

\subsection{3T3 Cells}
\begin{figure*}[!t]
    \centering
	\includegraphics[width=\textwidth]{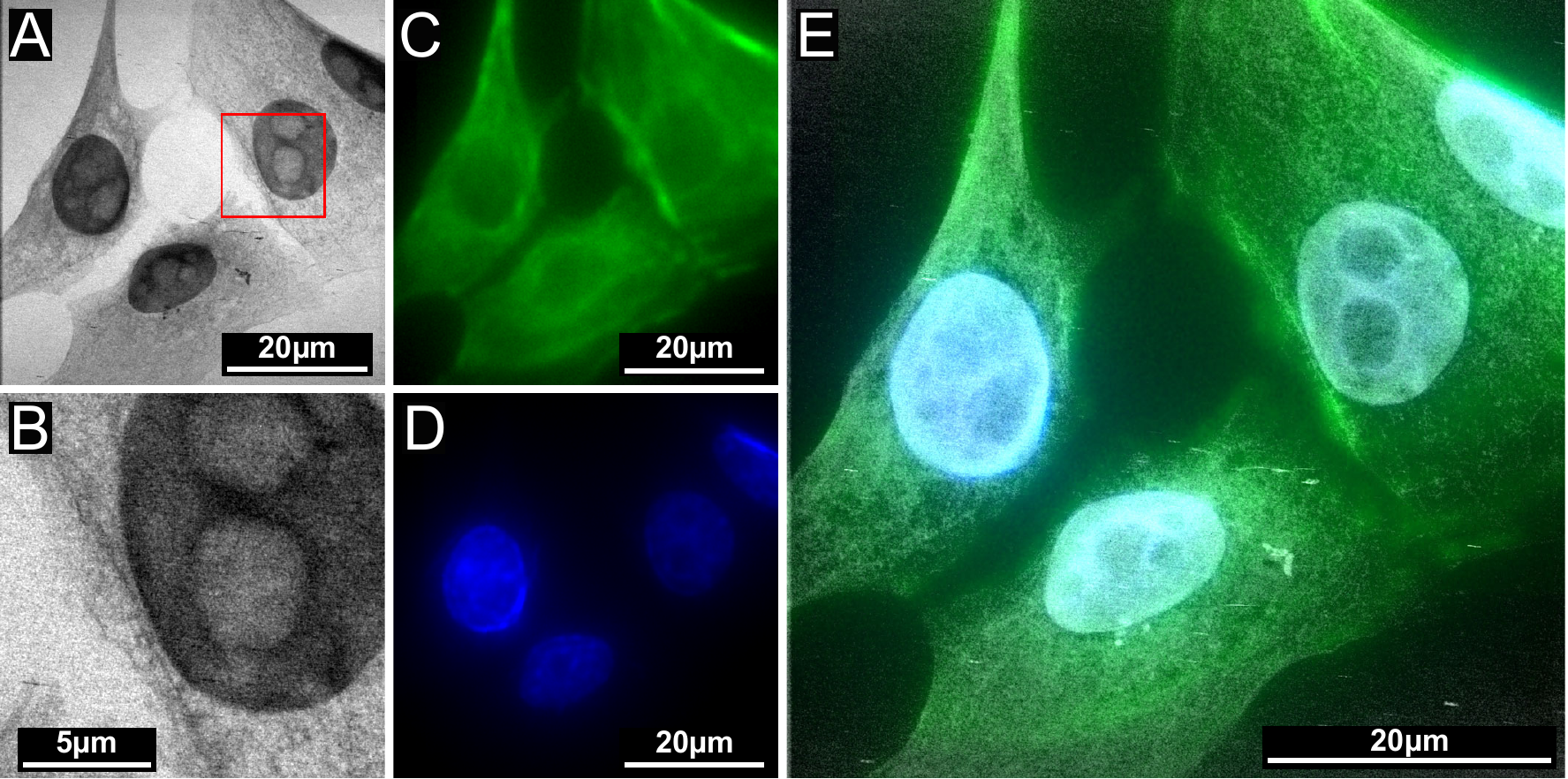}
	\caption[3T3 cells correlative]{Results of the correlative measurements of 3T3 cells. \textbf{A)} 70\,$\times$\,45\,s SXR image. 4 nuclei and the cytoskeleton are presented. \textbf{B)} Enlargement of the red-framed area of panel A. \textbf{C)} Actin sceleton measured with the green filter set. \textbf{D)} Nuclei measured with the blue filter set. \textbf{E)} Composite image of fluorescence images and SXR image. SXR contrast was inverted for better visibility of the fluorescence image.}
	\label{fig:3t3}
\end{figure*}

Next, we studied a conventional 3T3 cell culture. It was prepared as described earlier in the paper. As highlighted before, FLM images may highlight regions of interest for subsequent SXR recordings or help assigning signatures in SXR images to specific cellular structures. Therefore, we here labeled different cellular structures such as actin cytoskeleton, nucleus or mitochondria with respective fluorescent markers, i.e. fluorescently tagged phalloidin, DAPI and Mitotracker. As the signal from the Mitotracker was too weak and inconclusive, it was excluded from the evaluation. Figure \ref{fig:3t3} shows an examplary region of the sample that includes four 3T3 cells. Figures A and B show the SXR image at different magnifications, whereas C shows the FLM image of the actin staining in green and D shows the DAPI staining in blue. In E, a composition of all 3 channels is shown, for which, however, the SXR contrast has been inverted to allow better visibility of the fluorescence.

In the SXR image (Fig. \ref{fig:3t3}.A), the four dark oval-shaped components are easily identified as cell nuclei, which give a dark contrast due to their high (carbon) density. This is confirmed in the fluorescence image by DAPI staining (Fig. \ref{fig:3t3}.D). Furthermore, nucleoli are recognizable in the cell nuclei. In the magnified SXR image (Fig. \ref{fig:3t3}.B), which shows the red-framed section from Fig.~\ref{fig:3t3}.A, the high resolution of SXR microscopy becomes even clearer. Two nucleoli in the nucleus and, in particular, the cytoskeleton surrounding the nucleus can be seen, revealing the dense fiber network. The labeled actin cytoskeleton is also shown in the green fluorescence image (Fig.~\ref{fig:3t3}.C). In the composite image (Fig.~\ref{fig:3t3}.E), the interaction of the different contrasts is again particularly clear. Again, the SXR contrast was inverted for this image to allow better visibility of the fluorescence. In addition to what is displayed here,  other components of the cytoskeleton such as microtubules or intermediate filaments could be stained as well. Also cytoskeleton-associated proteins could be labeled.

\subsection{COS-7 cells}
\begin{figure*}[!t]
    \centering
	\includegraphics[width=\textwidth]{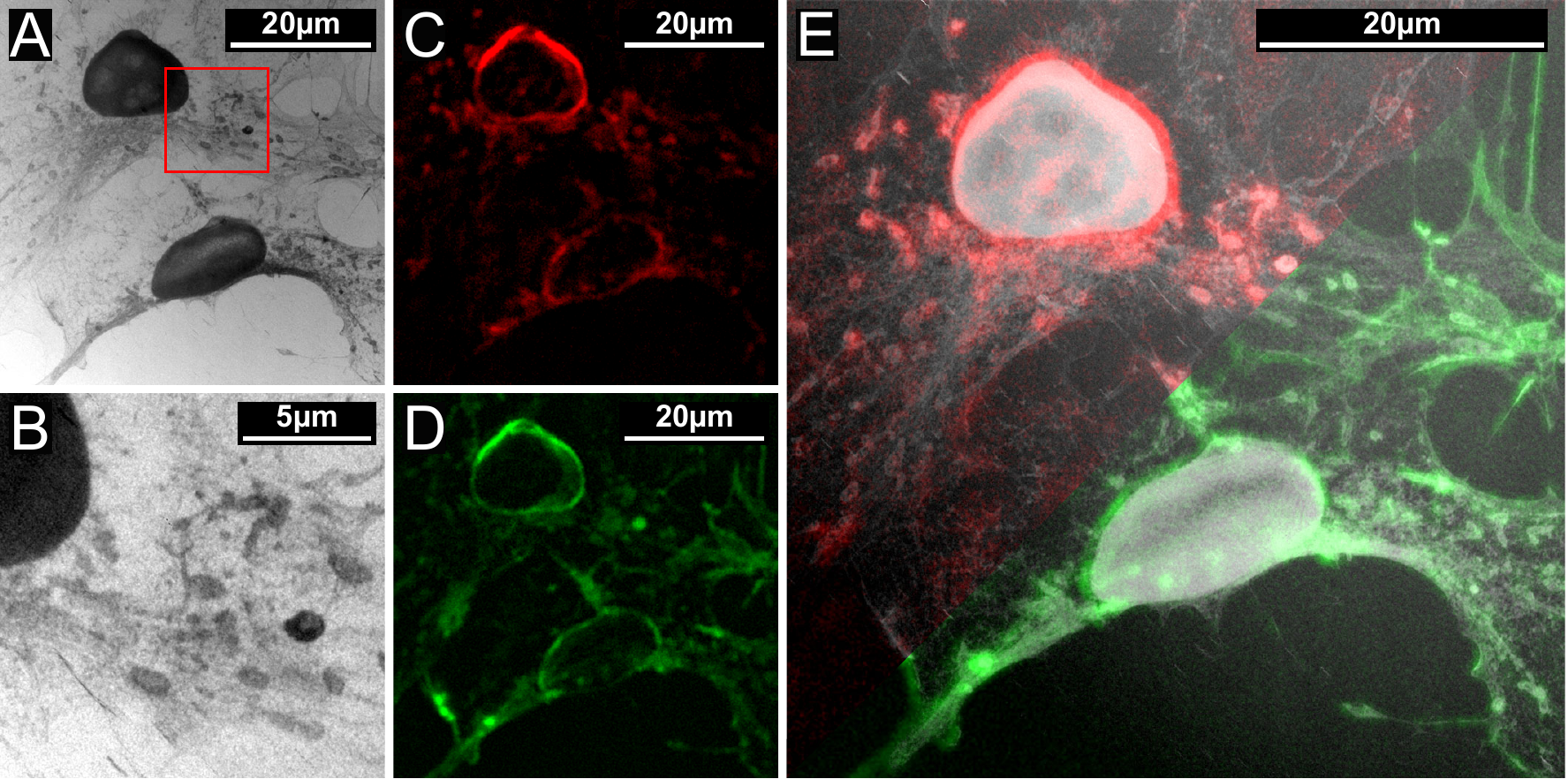}
	\caption[COS-7 cells correlative]{Results of the correlative measurements of COS-7 cells. \textbf{A)} 81\,$\times$\,45\,s SXR image. 2 nuclei and the surrounding cytoskeleton and mitochondria can be seen. \textbf{B)} Enlargement of the red-framed area of panel A. \textbf{C)} Mitochondria detected with the red filter set. \textbf{D)} Actin detected with the green filter set. \textbf{E)} Composite image of fluorescence images and SXR image. The SXR contrast was inverted for better visibility of the fluorescence image.}
	\label{fig:COS7}
\end{figure*}
The same procedure as for the 3T3 cells was performed for the COS-7 cells, except for labeling with DAPI. The recorded images are shown in Figure \ref{fig:COS7}. Panels A and B show the SXR image with different magnifications. FLM images are presented in C and D, whereas C shows the red channel with mitochondrial staining and D shows the green channel again with actin staining. In E a composite image is presented in split view. The upper left half shows the mitotracker and SXR, while the lower right half shows the actin staining and SXR. Again, the SXR contrast is inverted.

Since the COS-7 are also fibroblasts, similar components of the cells can be seen as in the 3T3 cells. These are again the nuclei and the cytoskeleton, but additionally the mitochondria, which were also fluorescently labeled. The nuclei seem to be thicker. Therefore, nucleoli are only (faintly) visible in the upper nucleus, see Fig.~\ref{fig:COS7}.A. Furthermore, the cytoskeleton is not as dense as in 3T3 cells, so that individual fibers can be detected. This effect is particularly evident in the composite image.(Fig. \ref{fig:COS7}.E). The mitochondria are visible as small particles distributed in the cytoskeleton, as shown in the enlarged section \ref{fig:COS7}.B. Identification is enabled by labeling with the mitotracker (Fig. \ref{fig:COS7}.C) and superposition of SXR and FLM images.

Both of these exemplary investigations of different cell types illustrate the interplay between structural SXR and functional FLM contrast. Especially labeling small organelles such as mitochondria and then being able to study them in the context of the whole cell visualized by SXR microscopy holds great potential. In a similar manner other organelles such as the Golgi apparatus or lysosomes can be stained as well.

\subsection{Photobleaching by SXR Irradiation}
\begin{figure}[h]
	\centering
	\includegraphics[width=0.55\linewidth]{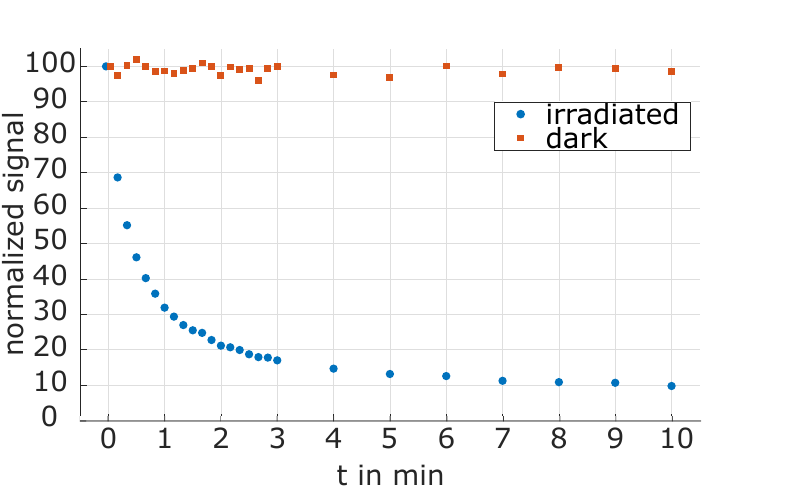}
	\caption[Qunatitative Fluorescence Measurement]{Quantitative measurement of the influence of SXR illumination of 200\,nm fluorescent nanobeads. The degradation of the fluorescence signal was normalized to the first data value to characterize the temporal behavior.}
	\label{fig:quantitative}
\end{figure}

Our setup also enables the investigation of the degradation of the fluorescence signal due to SXR irradiation. To this end, fluorescent particles with a diameter of 200\,nm were prepared on a SiN membrane and a measurement was performed as described in the methods section. In the results presented in Figure \ref{fig:quantitative}, the fluorescence signal of 200\,nm red fluorescent nanobeads was measured during constant irradiation (on the membrane) and without irradiation (on the chip) over a period of 10 minutes. The fluorescence signal was normalized to the first data value in order to compare its temporal evolution. The irradiated part of the sample shows a strong decrease in the detected signal compared to the dark part, where the fluorescence signal stays nearly constant. The behavior of the non-irradiated part can also be used to estimate the accuracy of this measurement and to rule out the possibility that the signal drop is caused by photobleaching from visible light. The strongest decay is observed during the first minute of irradiation. Presumably, the X-rays ionize the fluorescent molecules, permanently destroying them. This observation is consistent with the results published in \citep{hagen2012correlative} and leads to the conclusion that the fluorescence images should be taken before the SXR images. In fact, this is the preferred order anyway, as the entire sample is first scanned with the FLM to quickly identify ROIs. Regardless of this, these results show that the setup enables the characterization of the X-ray resilience of different fluorescent dyes. This is important with respect to the further development of correlative experiments.

\section{Conclusions}
The microscope presented here combines laboratory-based water-window X-ray microscopy and fluorescence microscopy in an integrated setup.
The light source for the X-ray microscope is based on a double-stream gas puff target with nitrogen as the target gas. It produces line emission at 431\,eV or 2.88\,nm. By using an ellipsoidal mirror as a condenser optic and a Fresnel zone plate as an objective lens, a resolution of 50\,nm half-pitch can be achieved, as demonstrated on a Siemens star test target. An algorithm based on cross-correlation has been developed to eliminate the effects of thermal drift on resolution.

The integration of a wide-field epifluorescence microscope into the SXR microscope allows correlation of fluorescence and X-ray images, i.e. correlation of functional and structural information. Specifically, FLM images can either be used to guide SXR recordings to regions of interest or assign signatures in the SXR images to specific structures. Direct integration of the FLM into the SXR microscope allows correlated image acquisition without moving the sample. This not only reduces the risk of sample destruction or alteration, but also significantly speeds up the measurement process by allowing relevant sample positions to be quickly identified and targeted for SXR recordings. Additionally, the FLM images can be used to assign signatures in the SXR images to specific structures. Multiple filter sets can be used for multi-color measurements, further enhancing the capabilities of the integrated setup. The result is a powerful tool for investigating different types of relevant life science samples, which was realized in a compact setup with a footprint of 1.5\,m\,$\times$\,4\,m.

As examples, fluorescent nanoparticles, cyanobacteria, and different types of cells, namely 3T3 and COS-7 cells with multi-color labeling, have been presented in this work. All samples were prepared on silicon nitride membranes and, in the case of the cells, were critical-point dried. Furthermore, quantitative measurements of fluorescence behavior under SXR irradiation are possible, which was demonstrated using nanobeads as an example. These studies can be of great benefit to all correlative X-ray microscopes, including those at synchrotrons.

Based on the high resolution and functional contrast demonstrated in this work, new milestones in lab-based correlative microscopy have come into reach. Advanced preparation methods could enable the examination of wet samples and full exploitation of the possibilities of the water window, e.g. by using microfluidic cells with SiN membranes as windows \citep{weinhausen2013microfluidic}. Progress in plasma source technology would lead to reduced exposure times and could enable tomography and cryofixation of the sample. Furthermore, future lensless imaging methods in the water window, enabled by coherent lab-based sources, can significantly profit by the achievements shown in this work.

\section{Competing Interests}
No competing interest is declared.

\section{Author Contributions Statement}
J.R., T.W. and S.F. conceived the microscope, J.R., S.K., J.A., F.W., Ma.Wü. and J.N. set up the microscope, J.R., S.K., J.A. and F.W. conducted the measurements, S.K., E.S., Ma.We, A.I. and F.H. prepared samples. J.R., S.K., G.G.P., and S.F. wrote the manuscript. All authors gave advice on different aspects of the microscope.

\section{Acknowledgments}
The authors thank Christian Rödel for providing the laser system,  Katharina Reglinski and Philipp Kellner for an introduction to nanoparticle preparation, Yashar Rouzbahani for preparing comparison samples, Christian Franke for discussions regarding data analysis, and Slawomir Skruszewicz for support with the gas puff target. We gratefully acknowledge funding by the ExNet initiative of the Helmholtz Association, Deutsche Forschungsgemeinschaft (PA 730/13), by NCN via 2020/39/I/ST7/03194, and by Laserlab Europe via Horizon 2020 (871124).

\bibliographystyle{abbrvnat}
\bibliography{Correlative_labbased_SXR_FLM_micro_arXiv_revised}

\end{document}